\pdfoutput=1

\documentclass[11pt,twoside,a4paper,cmspaper,final,collab]{cms-tdr}
\def\svnVersion{265423}\def\cmsCernNo{2013-037}\def\cmsCernDate{\today}\def\cmsMessage{Published in Physics Letters B as \href{http://dx.doi.org/10.1016/j.physletb.2014.10.032}{\doi{10.1016/j.physletb.2014.10.032.}}}

\begin{document}\cmsNoteHeader{TOP-12-026}

\hyphenation{had-ron-i-za-tion}
\hyphenation{cal-or-i-me-ter}
\hyphenation{de-vices}
\RCS$HeadURL: svn+ssh://svn.cern.ch/reps/tdr2/papers/TOP-12-026/trunk/TOP-12-026.tex $
\RCS$Id: TOP-12-026.tex 263518 2014-10-10 06:48:11Z fnguyen $
\newlength\cmsFigWidth
\ifthenelse{\boolean{cms@external}}{\setlength\cmsFigWidth{0.85\columnwidth}}{\setlength\cmsFigWidth{0.4\textwidth}}
\ifthenelse{\boolean{cms@external}}{\providecommand{\cmsLeft}{top\xspace}}{\providecommand{\cmsLeft}{left\xspace}}
\ifthenelse{\boolean{cms@external}}{\providecommand{\cmsRight}{bottom\xspace}}{\providecommand{\cmsRight}{right\xspace}}
\providecommand{\tauh}{\ensuremath{\tau_\mathrm{h}}\xspace}
\cmsNoteHeader{TOP-12-026}
\title{Measurement of the \ttbar production cross section in pp collisions at $\sqrt{s}=8$\TeV in dilepton final states containing one $\tau$ lepton}

\date{\today}

\abstract{
The top-quark pair production cross section is measured in final states
with one electron or muon and one hadronically decaying $\tau$ lepton
from the process
$\ttbar\to (\ell\nu_\ell) (\tau\nu_\tau) {\bbbar}$, where $\ell = \Pe,\mu$.
The data sample corresponds to an integrated luminosity of 19.6\fbinv
collected with the CMS detector
in proton-proton collisions at $\sqrt{s}=8$\TeV.
The measured cross section
$\sigma_{\ttbar} = 257\pm 3\stat \pm 24\syst \pm 7\lum\unit{pb}$,
assuming a top-quark mass of 172.5\GeV,
is consistent with the standard model prediction.
}

\hypersetup{%
pdfauthor={CMS Collaboration},
pdftitle={Measurement of the ttbar production cross section in pp collisions at sqrt(s)=8 TeV in dilepton final states containing one tau lepton},%
pdfsubject={CMS},%
pdfkeywords={CMS, physics, top quark, tau}}

\maketitle

\section{Introduction}

Top quarks at the CERN LHC
are mostly produced in pairs with the subsequent decays $\ttbar\to \PWp\cPqb\PWm\cPaqb$.
The decay modes of the two W bosons determine the event signature.
The dilepton decay channel corresponds to the case in which both W bosons decay into leptons,
where the term lepton usually refers to electrons or muons, as studied in Refs.~\cite{Aad:2014kva,Chatrchyan:2013faa}.
In this letter we measure the production cross section of top-quark pairs
by considering dilepton decays where one W boson promptly decays into $\ell\nu_\ell$,
with $\ell= \Pe$ or $\mu$, and the
other decays into $\tau\nu_\tau$,
$\ttbar\to (\ell\nu_\ell) (\tau\nu_\tau) \bbbar$.
The expected fraction of these events is 4/81 of all \ttbar decays.
The $\tau$ lepton is identified by means of its hadronic decay products,
with a branching fraction $\mathcal{B}(\tau\to \text{hadrons} + \nu_\tau) \simeq 65\%$,
to produce a narrow jet with a small number of charged hadrons,
denoted as $\tauh$.
The cross section is measured by counting the number of $\ell\tauh+X$ events consistent
with originating
from \ttbar production, after subtracting the contributions from other processes,
and correcting for the efficiency of the event selection.
A similar method was used in pp collisions at a centre-of-mass energy of
$\sqrt{s}=7$\TeV~\cite{Chatrchyan:2012vs}.
This ``$\tau$ dilepton'' channel is of particular interest because it is a natural background process
to the search for a charged Higgs
boson~\cite{Gunion:1989we,Djouadi:2005gj} with a mass smaller than that of the top quark.
In this case, the production chain $\ttbar\to\PH^+\cPqb\PWm\cPaqb$, with $\PH^+\to\tau^+\nu_\tau$
(or the corresponding charge-conjugate particles)
could give rise to differences with respect to the standard model (SM) prediction of the
number of $\ttbar$ events with a $\tau$ lepton~\cite{Chatrchyan:2012vca}.
The present measurement is based on data collected by the CMS experiment
in pp collisions at $\sqrt{s}=8$\TeV corresponding to an integrated luminosity of 19.6\fbinv.
The relative accuracy of this measurement improves over previous results~\cite{c:taudil_cdf1,c:taudil_cdf2,c:d0taudil,Aaltonen:2014hua,Aad:2012mza},
thanks to the inclusion of additional data and improved analysis techniques.

The CMS detector is briefly introduced in Section~\ref{sec:detector},
followed by details of the simulated samples in Section~\ref{sec:simulation},
and a brief description of the event reconstruction and event selection in Section~\ref{sec:eventsel}.
The descriptions of the background determination and the systematic uncertainties are given in
Sections~\ref{sec:background} and~\ref{sec:systematics}, respectively.
The measurement of the cross section is discussed in Section~\ref{sec:xsec}, and the
results are summarised in Section~\ref{sec:summ}.

\section{The CMS detector}
\label{sec:detector}

The central feature of the CMS apparatus is a superconducting solenoid
of 6\unit{m} internal diameter and 13\unit{m} in length, providing a magnetic field of 3.8\unit{T}.
Within the superconducting solenoid volume are a silicon pixel and strip tracker,
a lead tungstate crystal electromagnetic calorimeter,
and a brass/scintillator hadron calorimeter,
each composed of a barrel and two endcap sections.
The calorimetry provides
high-resolution energy and direction measurements of electrons and hadronic jets.
Muons are identified using
gas-ionization detectors embedded in the steel flux-return yoke outside the solenoid.
Extensive forward calorimetry complements the coverage provided by the barrel and endcap detectors.
The CMS experiment uses a right-handed coordinate system, with the origin at the nominal interaction point,
the $x$ axis pointing to the centre of the LHC ring, the $y$ axis pointing up (perpendicular to the LHC plane),
and the $z$ axis along the anticlockwise-beam direction.
The polar angle $\theta$ is measured from the positive $z$ axis and the azimuthal angle $\varphi$ is measured in the $x$-$y$ plane.
Charged particle trajectories are
measured covering $0 < \varphi \leq 2\pi$
in azimuth and $\abs{\eta}<2.5$, where the pseudorapidity $\eta$ is
defined as $\eta =-\ln[\tan ({\theta/2}) ]$.
The detector is nearly hermetic, allowing for energy balance
measurements in the plane transverse to the beam directions.
A two-level trigger system selects the most interesting proton-proton collision events for use in physics analyses.
A more detailed description of the CMS detector can be found elsewhere~\cite{JINST}.

\section{Data and simulation samples}
\label{sec:simulation}

Events are selected online by a trigger requiring a single isolated electron (muon) with transverse momentum $\pt>27\,(24)$\GeV
and $\abs{\eta}<2.5\,(2.1)$.

This measurement makes use of simulated samples of \ttbar events as well as other processes that mimic
the $\ell\tauh$ decay signature.
These samples are used to optimise the event selection, to calculate the acceptance for \ttbar events, and to estimate some of the backgrounds in the analysis.

The signal acceptance and \ttbar dilepton background are evaluated using a version of
\MADGRAPH which includes the effects of spin correlations~\cite{Alwall:2011uj,Alwall:2014hca}.
The number of expected \ttbar events is estimated with the next-to-next-to-leading-order (NNLO) SM cross section
of $251.7^{+6.3}_{-8.6}\,\text{(scale)} \pm6.5\,\mathrm{(PDF)}$\unit{pb}~\cite{Czakon:2013goa,Czakon:2011xx,Botje:2011sn,Gao:2013xoa,Ball:2012cx}
for a top-quark mass of 172.5\GeV,
where the first uncertainty is due to renormalisation and factorisation scales, and the second is due to the
choice of parton distribution functions (PDFs).
The generated events are subsequently processed with \PYTHIA 6.426~\cite{pythia}
which performs the hadronisation of partons. Soft radiation is matched
to the contributions from direct emissions accounted for in the matrix-element calculations
using the \kt-MLM approach~\cite{Alwall:2007fs}.
The $\tau$ lepton decays are simulated using \TAUOLA 27.121.5~\cite{tauola2},
which accounts for the $\tau$-lepton polarization.

The samples containing $\PW$+jet and $\cPZ$+jet events
are simulated using the \MADGRAPH 5.1.3.30 event generator~\cite{Artoisenet:2012st}.
The electroweak production of single top quarks is considered as a background process and is simulated with
\POWHEG 1.0, r1380~\cite{Nason:2004rx,powheg,Alioli:2010xd,Alioli:2009je,Re:2010bp}.
The diboson production processes $\PW\PW$, $\PW\cPZ$, and $\cPZ\cPZ$ are generated with \PYTHIA 6.424.
In each case, the \PYTHIA parameters for the underlying event are set according to the Z2* tune~\cite{Chatrchyan:2011id},
which uses the CTEQ6L PDFs~\cite{pdfset}.

Simulated events are processed using the full CMS detector simulation based on \GEANTfour~\cite{Agostinelli:2002hh,Allison:2006ve},
followed by a detailed trigger emulation and event reconstruction.
For both signal and background events, additional pp interactions (pileup) in the same or nearby bunch crossings are
simulated with \PYTHIA\ and superimposed on the hard collision, using a pileup multiplicity distribution
that reflects the luminosity profile of the analysed data.

\section{Event selection}
\label{sec:eventsel}

Events are reconstructed with the particle-flow (PF) algorithm~\cite{CMS-PAS-PFT-09-001,CMS-PAS-PFT-10-001},
which combines information from all sub-detectors to identify and reconstruct individual electrons, muons, photons,
charged and neutral hadrons.
The primary collision vertex is chosen as the reconstructed vertex with the largest $\sum\pt^2$ of the associated tracks.
Electrons are identified with a multivariate discriminant combining several quantities describing the track quality,
the shape of the energy deposits in the electromagnetic calorimeter, and the compatibility of the measurements from the tracker and the
electromagnetic calorimeter~\cite{CMS-PAS-EGM-10-004}, and are reconstructed with an average efficiency of approximately 95\%.
Muons are identified with additional requirements on the quality of the track reconstruction and on the number of measurements in the tracker
and the muon systems~\cite{Chatrchyan:2012xi}, and are reconstructed with an average efficiency of approximately 96\%.
Charged and neutral particles
provide the input to the anti-\kt jet clustering algorithm
with a distance parameter of 0.5~\cite{antikt}.
The jet momentum is determined from the vector sum of particle momenta in the jet.
After jet energies are corrected
for additional pileup contributions and for detector effects,
they are found in simulations to be within 5--10\% of the actual jet momentum~\cite{jme-10-011}.
The missing transverse energy \MET is calculated as the magnitude
of the vector sum of momenta from all reconstructed particles in the plane transverse to the beam.

In addition, higher-level observables such as
b-tagging discriminators and lepton isolation variables are used.
The lepton relative isolation is defined as
the transverse energy contributions deposited by charged hadrons ($E_\text{T, ch}$), neutral hadrons ($E_\text{T, nh}$),
and photons ($E_\text{T, ph}$) in a cone of radius $R=\sqrt{\smash[b]{(\Delta\varphi)^2 +(\Delta\eta)^2}}=0.4$
centered on the lepton candidate track,
relative to the lepton's transverse momentum ($\pt$), $I_\text{rel} = (E_\text{T, ch}+ E_\text{T, nh}+ E_\text{T, ph})/\pt$.
An electron (muon) candidate is considered to be non-isolated and is rejected if $I_\text{rel} > 0.1$ (${>}0.12$).

The hadronic products of the $\tau$-lepton decay are reconstructed using a
jet as the initial seed, and are then classified as having one or three charged hadrons
with the ``hadron-plus-strips"
algorithm~\cite{Chatrchyan:2012zz,Chatrchyan:2014nva}.
In the ``hadron-plus-strips" algorithm,
calorimeter energy deposits clustered along strips in the $\varphi$ direction are used for neutral pion identification.
Then, the decay modes, four-momenta, and isolation quantities of the $\tauh$ are determined, and
the following categories are considered: single hadron, hadron plus a strip, hadron plus two strips, and three hadrons.
These categories together encompass approximately 95\% of hadronic $\tau$-lepton decays.
The sum of the charged hadron charges
provides the $\tauh$ charge.
The $\tauh$-jet momentum is required to match the direction of the
original jet within a maximum distance $R = 0.1$.
Isolation criteria require that there be no additional charged hadrons with $\pt > 1.0$\GeV or photons with
transverse energy $\ET > 1.5$\GeV
within a cone
of size $R = 0.5$ around the direction of the $\tauh$ jet.
Electrons and muons misidentified as $\tauh$
are suppressed using
algorithms that combine information from the tracker, calorimeters, and muon detectors~\cite{JINST}.
The $\tauh$ identification efficiency is
defined as the ratio of the number of selected $\tauh$
candidates divided by the number of hadronic $\tau$-lepton decays in \ttbar events;
the ratio depends on \pt and $\eta$ of the $\tauh$, and is on average 50\%
for $\pt^{\tauh}>20$\GeV,
with a probability of approximately 1\% for
generic jets to be misidentified as a $\tauh$ jet.

The combined secondary vertex (CSV) algorithm~\cite{Chatrchyan:2012jua} is used to identify jets originating
from the hadronisation of b quarks. The algorithm combines the information about track impact parameters
and secondary vertices within jets into a likelihood discriminant to provide separation between b jets and
jets originating from light quarks, gluons, or charm quarks. The output of this CSV discriminant
has values between zero and one; a jet with a CSV value above a certain threshold is referred to as being ``b tagged''.
We choose a working point where the b-tagging efficiency is approximately 60\%,
as measured in a data sample of events enriched with jets from semileptonic b-hadron decays.
The misidentification rate of light-flavour jets is estimated from inclusive jet studies
and is measured to be about 0.1\% for jets with $\pt>30$\GeV.

Events are preselected by requiring
exactly one isolated electron (muon)
with transverse momentum $\pt>35\,(30)$\GeV and $\abs{\eta}<2.5\,(2.1)$, at least two jets with
$\pt>30$\GeV, and one additional jet with $\pt>20$\GeV.
The selected jets must be within $\abs{\eta}<2.4$. The electron or muon is required to be separated from any jet in the
($\eta, \varphi$) plane by a distance $R>0.4$. Events with any additional
loosely isolated, $I_\text{rel}<0.2$, electron (muon) of $\pt>15\,(10)$\GeV are rejected.
Further event selection requirements include
\MET$>40$\GeV and only one
$\tauh$ with $\pt> 20$\GeV and $\abs{\eta}<2.4$. The $\tauh$ and the lepton are required to have electric charges of opposite sign (OS).
At least one of the jets is required to be identified as originating from b-quark hadronisation (b-tagged).
\begin{figure}[htbp]
\centering
\includegraphics[width=0.48\textwidth]{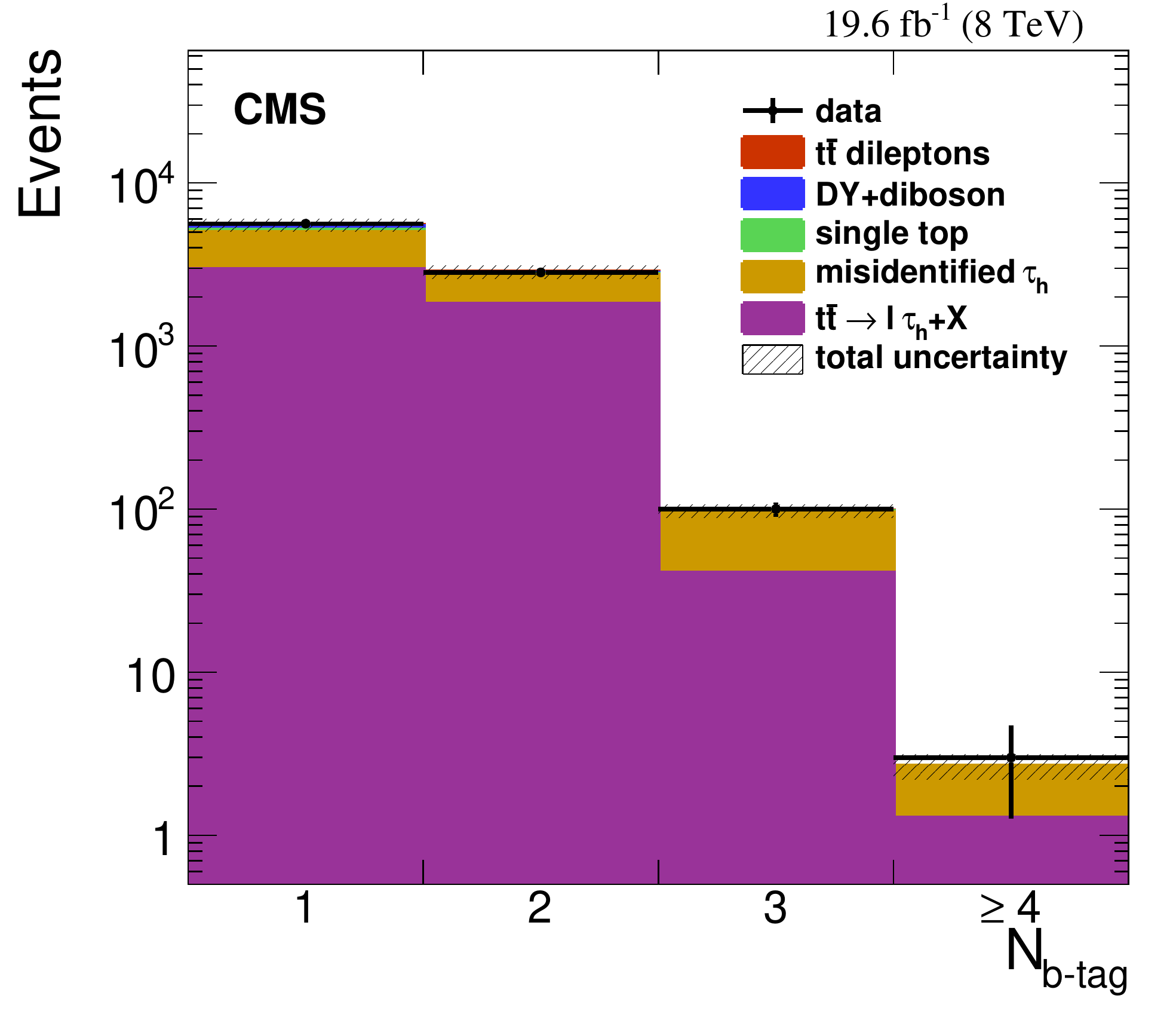}
\caption{
The \cPqb-tagged jet multiplicity after the full event selection.
The simulated contributions are normalised to the SM predicted values. The hatched area shows the total uncertainty.
}
\label{fig:btagmultiplicity}
\end{figure}

\begin{figure}[htbp]
\centering
\includegraphics[width=0.48\textwidth]{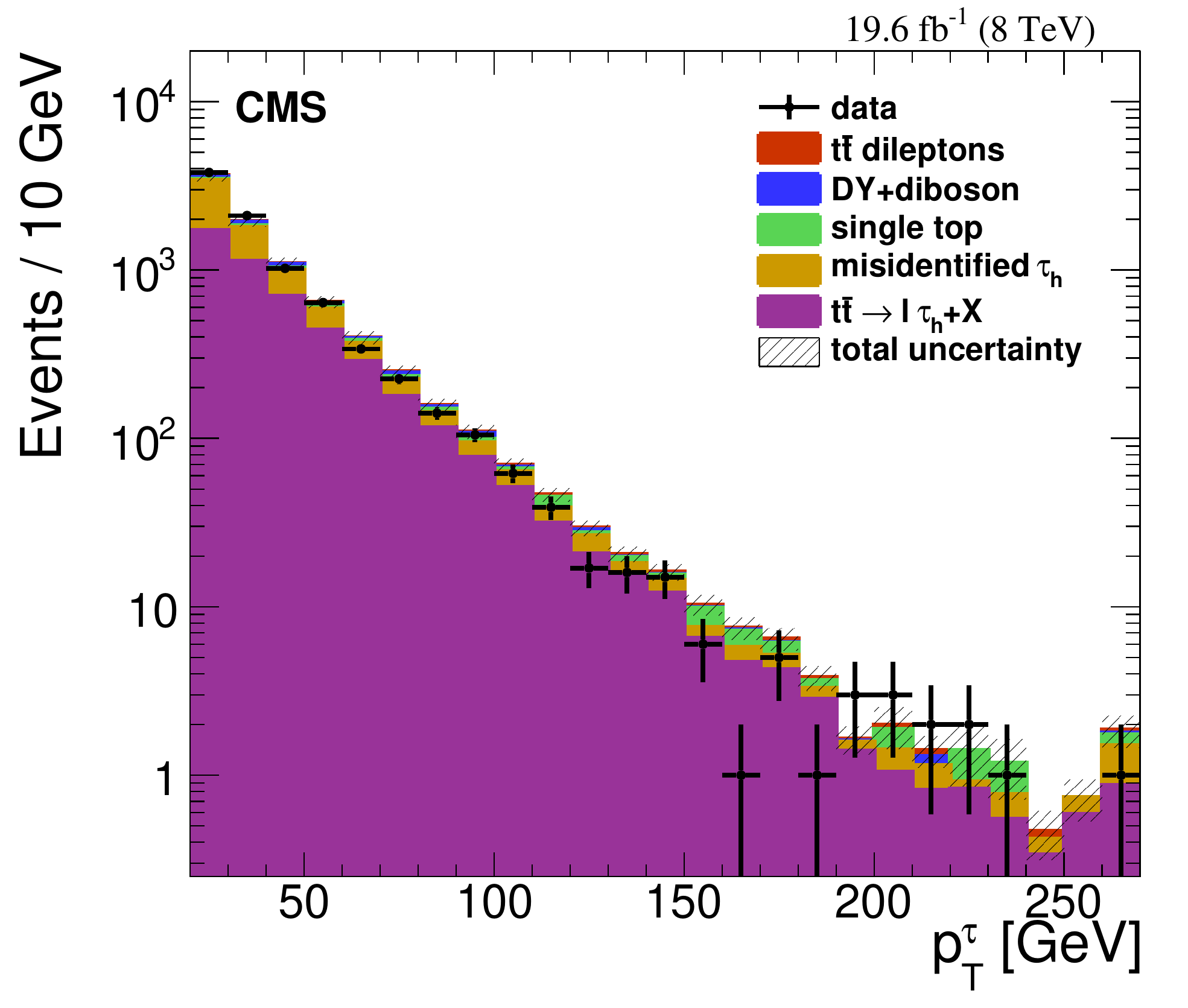}
\includegraphics[width=0.485\textwidth]{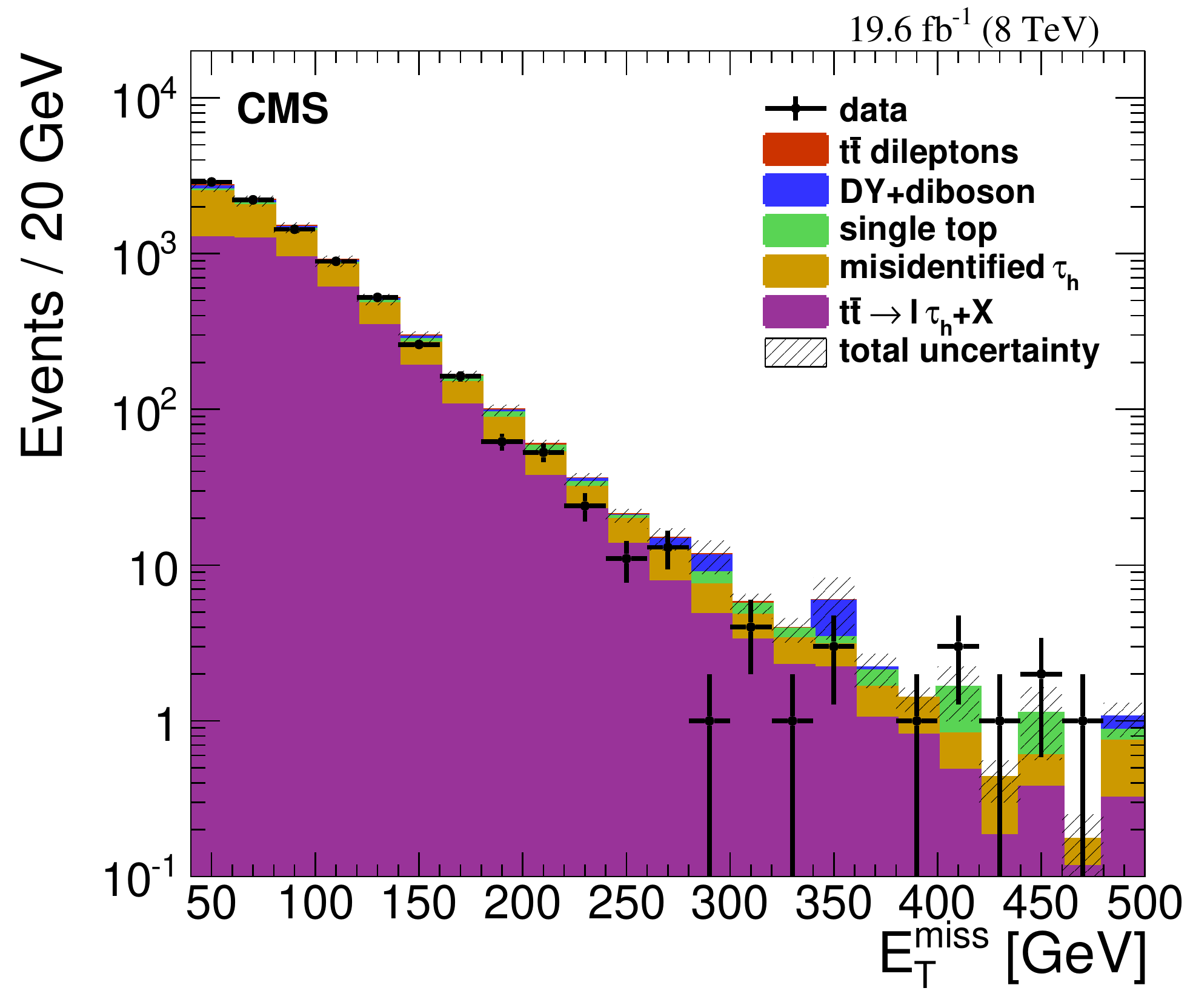}
\caption{
Distribution of the $\tauh$ \pt (\cmsLeft) and \MET (\cmsRight) after the full event selection,
for the $\Pe\tauh$ and $\mu\tauh$ channels combined.
The simulated contributions are normalised to the SM predicted values.
The hatched area shows the total uncertainty. The last bins include the overflow events.
}
\label{fig:taumetdistribution}
\end{figure}

Figure~\ref{fig:btagmultiplicity} shows,
for the sum of the $\Pe\tauh$ and $\mu\tauh$ final states,
a comparison between data and simulation of the number of b-tagged jets in each
event $N_\text{b-tag}$ after all the selection criteria have been applied.
The distributions of the $\tauh$ \pt and \MET after the final event selection
are shown in the \cmsLeft and \cmsRight panels of Fig.~\ref{fig:taumetdistribution}, respectively.
The distributions show agreement between the observed numbers of events and the expected
numbers of signal and background events obtained from the
simulated distributions normalised to the integrated luminosity of the selected data sample.

\begin{figure}[htbp]
\centering
\includegraphics[width=0.48\textwidth]{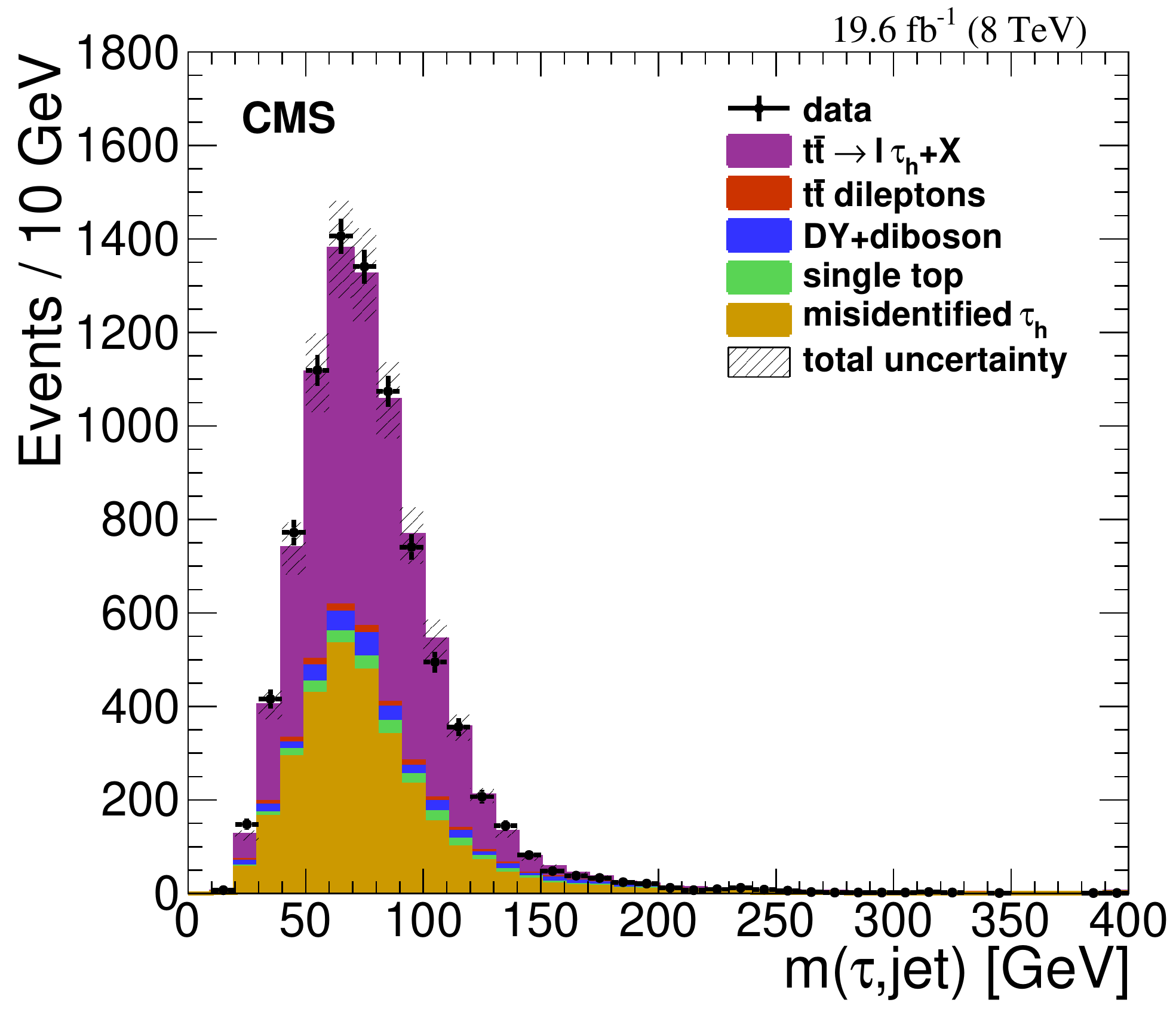}
\includegraphics[width=0.485\textwidth]{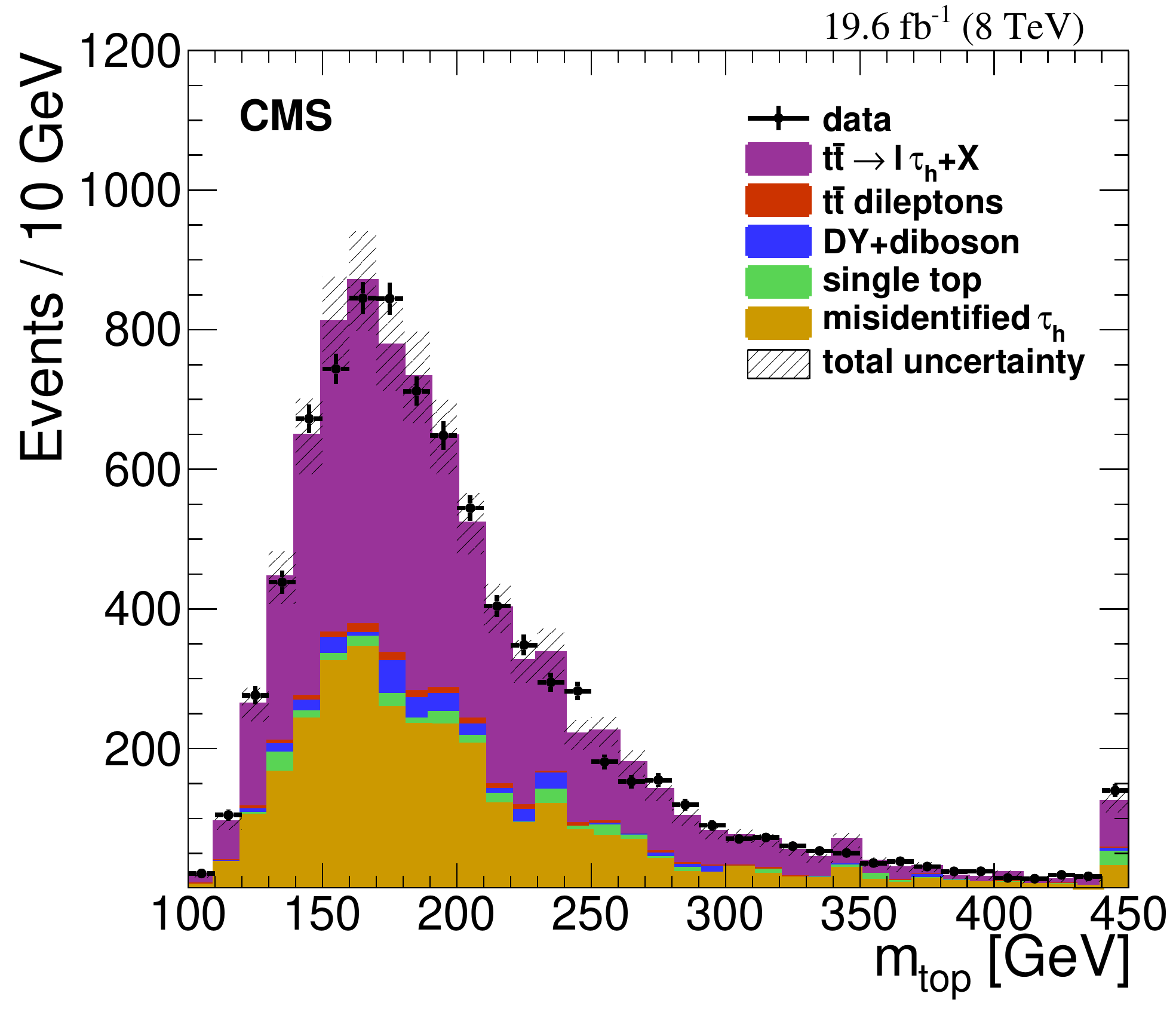}
\caption{(\cmsLeft)
Minimum invariant mass reconstructed by pairing the $\tauh$ with either a b-tagged jet or with the highest \pt non b-tagged jet, as described in the text.
(\cmsRight) Distribution of the reconstructed top-quark mass $m_\text{top}$ for the $\ell\tauh$ candidate events after the full event selection.
Data (points) are compared with the sum of signal and background yields,
for the $\Pe\tauh$ and $\mu\tauh$ channels combined. The simulated contributions are normalised to the SM predicted values.
The hatched area shows the total uncertainty. The last bins include the overflow events.
}
\label{fig:taudileptonmtop}
\end{figure}

Following the final selection, additional kinematic features of the \ttbar events are studied
to evaluate the agreement between the observed data and the predicted
sum of signal and background.
For each event, two invariant mass combinations are reconstructed by pairing the $\tauh$
with the two candidate b-jets:
(1)~in events with two or more b-tagged jets, the two combinations are based on the two
     b-tagged jets with the highest value of the discriminator;
(2)~in events
with one b-tagged jet,
this is used for the first combination,
while the non-b-tagged jet with the highest \pt is used to form the second combination.
For the two combinations, the invariant mass with the lowest value is shown in Fig.~\ref{fig:taudileptonmtop} (left),
for the $\Pe\tauh$ and $\mu\tauh$ channels combined.

For each event, the top-quark mass $m_\text{top}$ is reconstructed using the KINb algorithm~\cite{Chatrchyan:2012ea,Chatrchyan:2011nb}.
Due to the multiple neutrinos in the event, the reconstruction of $m_\text{top}$ leads to an underconstrained system.
The KINb algorithm applies constraints on the W boson mass, the mass difference between the top and anti-top quark,
and the longitudinal momentum of the \ttbar system.
For each event, solutions to the kinematic equations are evaluated, varying the jet momenta and the direction of \MET
within their resolutions.
For each set of variations and each lepton-jet combination, the kinematic equations allow up to four solutions; the one with the
lowest \ttbar invariant mass is accepted if the mass difference between the two top quarks is less than 3\GeV.
For each event, the accepted solutions corresponding to the two possible lepton-jet combinations are counted and the
combination with the largest number of solutions is chosen and $m_\text{top}$ is obtained by fitting the peak of this distribution.
The events in which solutions are found
are shown in Fig.~\ref{fig:taudileptonmtop} (right).
Data are in agreement with the expected sum of signal and background events.

\section{Background estimate}
\label{sec:background}

The main background (misidentified $\tauh$) comes from events with one lepton (electron or muon), significant \MET, and three or more jets,
where one jet is misidentified as a $\tauh$ jet~\cite{Chatrchyan:2012vca}.
The dominant source is \ttbar lepton+jet events.
The misidentified $\tauh$ background accounts also for events with
W bosons produced in association with jets, either genuine W+jet or single-top-quark production,
and for QCD multijet events.
In order to estimate this background from data, the misidentification probability $w(\text{jet} \to \tauh)$
is parameterised as a function of the jet \pt, $\eta$, and width ($R_\text{jet}$).
The quantity $R_\text{jet}$ is defined as $\sqrt{\smash[b]{\sigma_\eta^2 + \sigma_\varphi^2 }}$, where
$\sigma_\eta$ ($\sigma_\varphi$) expresses the extent in $\eta$ ($\varphi$) of the jet cluster~\cite{jme-10-011}.

The probability $w(\text{jet} \to \tauh)$ is evaluated from two control samples:
\begin{itemize}
\item $w_{W+\text{jets}}$: from a W+jet event sample, selected by requiring one isolated muon with $\pt > 20$\GeV
and $\abs{\eta} < 2.1$, and at least one jet with $\pt > 20$\GeV and $\abs{\eta} < 2.4$;
\item $w_\mathrm{QCD}$: from a QCD multijet sample, triggered by one jet with $\pt >40$\GeV, selected
by requiring events to have at least two jets with $\pt >20$\GeV and $\abs{\eta} < 2.4$,
where the triggering jet is removed from the misidentification rate calculation to avoid a trigger bias.
\end{itemize}
Both probabilities are evaluated in simulated events as well as in data,
with good agreement found between the results from simulation and data~\cite{Chatrchyan:2012zz}.

The number of events containing misidentified $\tauh$ candidates is then determined as

\begin{equation}
\label{eq:fakes}
N^\text{misid} = \sum^{M}_{i}~ \sum^{m}_{j} w_{i}^{j} (\text{jet} \to\Pgt) - N^\text{other},
\end{equation}

\noindent
where $j$ is the jet index of event $i$, and $m$ is the number of jets in each event and $M$ is the total number of events.
The quantity $N^\text{other}$ is the expected ${\simeq}$20\% contamination from signal and other processes
to the misidentified background as estimated from simulated samples.
The value of $N^\text{other}$ is evaluated by applying the procedure described above to simulated events of
$\cPZ/\gamma^{\ast} \to \tau \tau$,
single-top-quark production, diboson production, and the \ttbar processes
included in the misidentified $\tauh$ background estimation.

Jets in QCD multijet events originate mainly from gluons, while in W+jet events they are predominantly from quarks.
The quark and gluon composition in the misidentified $\tauh$ events lies between these two control samples.
As $w_\mathrm{QCD}<w_{\PW+\text{jets}}$,
the actual $N^\text{misid}$ value is under-
(over\hbox{-}\nobreak\hskip0pt)
estimated by applying the $w_\mathrm{QCD}$ ($w_{\PW+\text{jets}}$) probability.
We determine from data the rate for the misidentification of a jet to be identified as a $\tauh$,
and from simulation the quark/gluon composition in the W+jet and multijet samples.
From these quantities we derive the following combination:

\begin{equation}
\label{eq:weighave}
\langle N^\text{misid}\rangle = SF_{\PW+\text{jet}}\times N^\text{misid}_{\PW+\text{jet}} + SF_\mathrm{QCD}\times N^\text{misid}_\mathrm{QCD},
\end{equation}

where the misidentification rates, extracted from the data control samples discussed above,
are combined with the scale factors $SF$s determined
from the set of equations describing the quark/gluon composition of the samples:
$SF_\mathrm{QCD}=0.83$ and $SF_{\PW+\text{jet}}=0.17$.
The corresponding systematic uncertainty is
obtained from Eq.~(\ref{eq:weighave}) by weighting the
relative
deviations of $N^\text{misid}_{\PW+\text{jet}}$ and
$N^\text{misid}_\mathrm{QCD}$ from $\langle N^\text{misid}\rangle$
with the related scale factors. This results in an uncertainty of 7\% for both $\Pe\tauh$ and $\mu\tauh$ channels.

The efficiency of the OS requirement $\varepsilon _\mathrm{OS}$ is
determined from simulated lepton+jet \ttbar events and is applied in order to obtain the misidentified $\tauh$ background
after the final event selection $N_\mathrm{OS}^\text{misid}$, where
$N_\mathrm{OS}^\text{misid} = \varepsilon_\mathrm{OS}\cdot N^\text{misid}$.
We find values of
$\varepsilon_\mathrm{OS}=0.729 \pm 0.002\stat \pm 0.004\syst$
for the $\Pe\tauh$ selection and
$\varepsilon_\mathrm{OS}=0.731 \pm 0.002\stat \pm 0.003\syst$
for the $\mu\tauh$ selection,
where all sources of systematic uncertainty are accounted for in
the modelling of the simulated \ttbar lepton+jet events.

\section{Systematic uncertainties}
\label{sec:systematics}

Several sources of systematic uncertainty are considered and listed in Table~\ref{tab:SummarySystematics}.
They are related both to the signal reconstruction efficiency, background determination, and
luminosity measurement (Experimental uncertainties) and to the theoretical assumptions on
the \ttbar production (Theoretical uncertainties).
In Table~\ref{tab:SummarySystematics} and in what follows, relative
values refer to the cross section uncertainty unless explicitly stated otherwise.
\begin{table*}[htpb]
\topcaption{List of systematic uncertainties in the cross section measurement, and their combination.
Lepton reconstruction uncertainties are uncorrelated,
while all other uncertainties are assumed 100\% correlated.
}
\label{tab:SummarySystematics}
\small
\setlength{\extrarowheight}{1.5pt}
\centering
\begin{tabular}{lccc}
\multicolumn{4}{c}{ } \\
\hline

\multicolumn{1}{c}{Source} & \multicolumn{3}{c}{Uncertainty [\%]} \\ \cline{2-4}
                                               &  $\Pe\tauh$        & $\mu \tauh$    &     Combined \\
\hline
\multicolumn{1}{l}\textit{Experimental uncertainties:} & & &\\
    $\tauh$ jet identification                            &       6.0         &       6.0     &   6.0\\
    $\tauh$ misidentification background                  &       4.3         &       4.3     &   4.3\\
    $\tauh$ energy scale                                  &       2.4         &       2.5     &   2.5\\
    b-jet tagging, jet misidentification                            &       1.6         &       1.6     &   1.6\\
    jet energy scale, jet energy resolution, \MET        &       1.9         &       1.9     &   1.9\\
    lepton reconstruction                                &       0.8         &       0.6     &   0.5\\
    other backgrounds                &       0.6         &       0.7     &   0.7\\
    luminosity                                           &       2.6         &       2.6     &   2.6\\ \hline
\multicolumn{1}{l}\textit{Theoretical uncertainties:} & & &\\
    matrix element-parton shower matching                &       1.7         &       1.3     &   1.5\\
    factorisation/renormalisation scale            &       2.9         &       2.9     &   2.9\\
    generator                                            &       1.5         &       1.5     &   1.5\\
    hadronisation                                        &       1.7         &       1.7     &   1.7\\
    top-quark \pt modelling                                    &       0.7         &       0.5     &   0.6\\
    parton distribution functions                   &       0.8         &       0.7     &   0.7\\
\hline
\multicolumn{1}{l}{total systematic uncertainty}       &      {9.6}         &      {9.5}     & {9.5}\\
\hline
\end{tabular}
\end{table*}

\subsubsection*{Experimental uncertainties}
Regarding the $\tauh$ reconstruction, the uncertainty associated
to the identification efficiency amounts to 6\%, while the contribution relative to
the $\tauh$ jet energy scale is 2.4\% (2.5\%) for the $\Pe\tauh$ ($\mu\tauh$) channel,
as estimated by varying the \pt of the $\tauh$
jet by 3\%~\cite{Chatrchyan:2012zz,Chatrchyan:2014nva}.
The uncertainty in the $\tauh$ identification efficiency includes the uncertainty in charge determination
which is estimated to be smaller than 1\%.
The uncertainty related to
the misidentified $\tauh$ background process, discussed in Section~\ref{sec:background},
is obtained by propagating the 7\% uncertainty on $\langle N^\text{misid}\rangle$ to the cross section
determination and results in 4.3\% for both channels. It also includes the uncertainty in the OS efficiency determination.

The reconstruction of a light flavour jet as a b quark is defined as mistagging.
The uncertainty due to b (mis)tagging
is estimated to reflect the data-to-simulation scale factors and corresponding
uncertainties for b-tagging and mistagging efficiencies~\cite{Chatrchyan:2012jua}. When propagated
to the cross section measurement, they amount to 1.6\% for both $\Pe\tauh$ and $\mu\tauh$ channels.

The jet energy scale (JES) uncertainty is estimated~\cite{jme-10-011}
by varying the jet energy within the \pt- and $\eta$-dependent JES uncertainties per jet,
and taking into account the uncertainty due to pileup and parton flavour.
The jet energy resolution (JER) is estimated by smearing the jet energy in simulation within the $\eta$-dependent JER uncertainties
per jet. The JES and JER uncertainties are propagated in order to estimate the uncertainty of the \MET scale.
In addition, modelling of the \MET component, which is not clustered in jets, is also considered.
The resulting uncertainty from propagating these effects to the cross section measurement is 1.9\% for both the
$\Pe\tauh$ and $\mu\tauh$ channels.

Uncertainties due to trigger, lepton identification, isolation, and lepton energy scale
are calculated from independent samples with a ``tag-and-probe'' method~\cite{CMS-PAS-EGM-10-004,Chatrchyan:2012xi}, and
yield 0.8\% (0.6\%) for the $\Pe\tauh$ ($\mu\tauh$) channel.

An overall 0.6\% (0.7\%) uncertainty for the $\Pe\tauh$ ($\mu\tauh$) channel
is due to other minor backgrounds,
accounting for the uncertainties related to
the theoretical cross sections, JES, and b-tagging in these simulated samples,
and the $\ell\to\tauh$ ($\ell=\Pe,\mu$) misidentification
in the $\cPZ/\gamma^\ast\to\ell^+\ell^-$ and \ttbar dilepton processes.

Finally, the integrated luminosity is known with 2.6\% accuracy~\cite{CMS-PAS-LUM-13-001}.

\subsubsection*{Theoretical uncertainties}
The theoretical uncertainty due to the matrix element (ME)
and parton shower (PS) matching is
estimated by varying up and down by a factor of two the threshold between jet production at the ME level and via PS,
and it results in 1.7\% (1.3\%) for the $\Pe\tauh$ ($\mu\tauh$) channel.

The modelling uncertainty in the signal acceptance
due to the factorisation and renormalisation scale choices is estimated by varying them simultaneously
up and down by a factor of two from the nominal value equal to the $Q^2$ in the event,
with an uncertainty of 2.9\% found for both channels.

The uncertainty due to the choice of the generator is estimated as the
relative difference between
the acceptances evaluated with
\MADGRAPH and \POWHEG \cite{Nason:2004rx,powheg,Alioli:2010xd,Alioli:2011as} after the full event selection and results in 1.5\%.
In a similar way, the uncertainty in the hadronisation scheme is evaluated from the
relative differences between
the acceptances from \POWHEG{+}\PYTHIA and \POWHEG{+}\HERWIG samples,
estimated prior to the b-tagging or $\tauh$ jet requirement, resulting in a 1.7\%
uncertainty.

We consider the uncertainty related to the top-quark \pt scale modelling by varying
the top-quark \pt spectrum and evaluating the change in the signal acceptance, resulting in 0.6\%,
and the uncertainty related to the PDF variations following the PDF4LHC prescriptions~\cite{Botje:2011sn}, resulting in 0.7\%.

\section{Cross section measurement}
\label{sec:xsec}

The number of expected signal and background events as well as the number of observed events after
all selections are summarised in Table~\ref{tab:SummaryEventYieldTauHadHPS}.
The statistical and
systematic uncertainties are also shown.
\begin{table*}[htpb]
\centering
\topcaption{
Number of expected events for signal (assuming $m_\text{top}=172.5$\GeV) and backgrounds.
The background from misidentified $\tauh$ is estimated from data, while the other backgrounds are estimated from simulation.
Statistical and systematic uncertainties are shown.
}
\label{tab:SummaryEventYieldTauHadHPS}
\setlength{\extrarowheight}{1.5pt}
\begin{tabular}{ccc}
\hline
\multicolumn{1}{c}{Source} & $\Pe\tauh$  &            $\mu\tauh$                          \\
\hline
misidentified $\tauh$                                          & 1341 $\pm$ 3 $\pm$ 94  &   1653 $\pm$ 3 $\pm$ 116    \\
$\ttbar \to (\ell \nu_\ell) (\ell \nu_\ell )\bbbar$ & 55  $\pm$ 1 $\pm$ 3    &   68  $\pm$ 2 $\pm$ 4 \\
$\cPZ/\gamma ^{\ast}\to \Pe\Pe,\mu\mu$                         & 11  $\pm$ 5 $\pm$ 5     &   12  $\pm$ 5  $\pm$ 5   \\
$\cPZ/\gamma ^{\ast}\to \tau\tau$                          & 85  $\pm$ 14 $\pm$ 8    &   166  $\pm$ 20  $\pm$ 18  \\
single top quark                                            & 104  $\pm$ 7 $\pm$ 9    &   133  $\pm$ 8  $\pm$ 10  \\
dibosons                                                        & 15  $\pm$ 1 $\pm$ 1     &   19  $\pm$ 1 $\pm$ 1   \\
\hline
total expected background                                     & 1611 $\pm$ 17 $\pm$ 95   &  2051 $\pm$ 22 $\pm$ 118  \\
\hline
expected signal yield
& 2134 $\pm$ 9 $\pm$ 170 &   2632 $\pm$ 11 $\pm$ 212  \\
\hline
data                                                            & 3779                    &   4767         \\
\hline
\end{tabular}
\end{table*}
The \ttbar production cross section measured from $\tau$ dilepton events is $\sigma_{\ttbar}= (N - B)/(L \cdot  A_\text{tot})$,
where $N$ is the number of observed candidate events, $B$ is the
estimate of the background and $L$ is the integrated luminosity.
The total acceptance $A_\text{tot}$ is
the product of the branching fractions, geometrical and kinematic acceptance, trigger,
lepton identification, and the overall reconstruction efficiency. It is evaluated with respect to the inclusive \ttbar sample.
After the OS requirement and assuming a top-quark mass $m_\text{top}=172.5$\GeV, we obtain:
\begin{equation*}\begin{split}
\label{eq:efficiency}
A_\text{tot} (\Pe\tauh) &= 0.04333\pm 0.00017\stat \pm 0.00300\syst\,\%;\\
A_\text{tot} (\mu\tauh) &= 0.05370\pm 0.00021\stat \pm 0.00376\syst\,\%.
\end{split}
\end{equation*}
The statistical uncertainties are due to the limited number of simulated events and the systematic uncertainties are estimated by accounting
for all sources listed in Table~\ref{tab:SummarySystematics}.
The statistical and systematic uncertainties listed in Table~\ref{tab:SummaryEventYieldTauHadHPS} are propagated
to the final cross section measurements:
\begin{equation*}\begin{split}
\label{eq:xsresult}
\sigma_{\ttbar} (\Pe\tauh) &= 255 \pm 4\stat \pm 24\syst \pm 7\lum\unit{pb}; \\
\sigma_{\ttbar} (\mu\tauh) &= 258 \pm 4\stat \pm 24\syst \pm 7\lum\unit{pb}.
\end{split}
\end{equation*}
The BLUE method~\cite{blue} is used to combine the cross section measurements in
the $\Pe\tauh$ and $\mu\tauh$ channels, yielding weights of 0.47 and 0.53, respectively.
Lepton reconstruction uncertainties are uncorrelated,
while all other uncertainties are assumed 100\% correlated. With this method we obtain a combined result of
$\sigma_{\ttbar}=257\pm 3\stat \pm 24\syst \pm 7\lum\unit{pb}$,
in agreement with the NNLO expectation of
$251.7\,^{+6.3}_{-8.6}\,(\text{scales}) \pm6.5\,(\mathrm{PDF})$\unit{pb}.
Following the most recent conventions for the treatment of PDF and scale uncertainties the same calculation yields
$252.9\,^{+6.4}_{-8.6}\,(\text{scale}) \pm11.7\,(\mathrm{PDF}+ \alpha_\mathrm{S})$\unit{pb}~\cite{Czakon:2013goa,Czakon:2011xx,Botje:2011sn,Gao:2013xoa,Ball:2012cx}.
The dependence on the top-quark mass has been studied for the range 160--185\GeV and is well described by a linear variation.
If we adjust our result to the current world average value of 173.3\GeV~\cite{ATLAS:2014wva}, we obtain a cross section
that is lower by 3.1\unit{pb}.

\section{Summary}
\label{sec:summ}
A measurement
of the \ttbar production cross section in the channel
$\ttbar\to (\ell\nu_\ell) (\tau \nu_\tau) \bbbar$
is presented,
where $\ell$ is an electron or a muon,
and the $\tau$ lepton is reconstructed through its hadronic decays.
The data sample corresponds to an integrated luminosity of 19.6\fbinv collected in proton-proton collisions at $\sqrt{s}=8$\TeV.
Events are selected by requiring the presence of one isolated electron or muon, two or more jets (at least one of which is b-tagged),
significant missing transverse energy, and one $\tau$.
The largest background contribution is estimated from data and consists of \ttbar events with one W boson decaying into jets,
where one jet is misidentified as a $\tau$.
The measured cross section
is $\sigma_{\ttbar} = 257\pm 3\stat \pm 24\syst \pm 7\lum\unit{pb}$
for a top-quark mass of 172.5\GeV.
This measurement improves over previous results in this decay channel,
and it is in good agreement with the standard model expectation and other measurements of the \ttbar cross section at same centre-of-mass energy.

\section*{Acknowledgements}

We congratulate our colleagues in the CERN accelerator departments for the excellent performance of the LHC and thank the technical and administrative staffs at CERN and at other CMS institutes for their contributions to the success of the CMS effort. In addition, we gratefully acknowledge the computing centres and personnel of the Worldwide LHC Computing Grid for delivering so effectively the computing infrastructure essential to our analyses. Finally, we acknowledge the enduring support for the construction and operation of the LHC and the CMS detector provided by the following funding agencies: BMWFW and FWF (Austria); FNRS and FWO (Belgium); CNPq, CAPES, FAPERJ, and FAPESP (Brazil); MES (Bulgaria); CERN; CAS, MoST, and NSFC (China); COLCIENCIAS (Colombia); MSES and CSF (Croatia); RPF (Cyprus); MoER, ERC IUT and ERDF (Estonia); Academy of Finland, MEC, and HIP (Finland); CEA and CNRS/IN2P3 (France); BMBF, DFG, and HGF (Germany); GSRT (Greece); OTKA and NIH (Hungary); DAE and DST (India); IPM (Iran); SFI (Ireland); INFN (Italy); NRF and WCU (Republic of Korea); LAS (Lithuania); MOE and UM (Malaysia); CINVESTAV, CONACYT, SEP, and UASLP-FAI (Mexico); MBIE (New Zealand); PAEC (Pakistan); MSHE and NSC (Poland); FCT (Portugal); JINR (Dubna); MON, RosAtom, RAS and RFBR (Russia); MESTD (Serbia); SEIDI and CPAN (Spain); Swiss Funding Agencies (Switzerland); MST (Taipei); ThEPCenter, IPST, STAR and NSTDA (Thailand); TUBITAK and TAEK (Turkey); NASU and SFFR (Ukraine); STFC (United Kingdom); DOE and NSF (USA).

Individuals have received support from the Marie-Curie programme and the European Research Council and EPLANET (European Union); the Leventis Foundation; the A. P. Sloan Foundation; the Alexander von Humboldt Foundation; the Belgian Federal Science Policy Office; the Fonds pour la Formation \`a la Recherche dans l'Industrie et dans l'Agriculture (FRIA-Belgium); the Agentschap voor Innovatie door Wetenschap en Technologie (IWT-Belgium); the Ministry of Education, Youth and Sports (MEYS) of the Czech Republic; the Council of Science and Industrial Research, India; the HOMING PLUS programme of Foundation for Polish Science, cofinanced from European Union, Regional Development Fund; the Compagnia di San Paolo (Torino); the Consorzio per la Fisica (Trieste); MIUR project 20108T4XTM (Italy); the Thalis and Aristeia programmes cofinanced by EU-ESF and the Greek NSRF; and the National Priorities Research Program by Qatar National Research Fund.

\bibliography{auto_generated}   

\providecommand{\href}[2]{#2}\begingroup\raggedright\begin{thebibliography}{10}%
\makeatletter
\providecommand{\hrefCMSnoop }[0]{\@secondoftwo}%
\makeatother
\providecommand{\doi}{\texttt{doi:}\begingroup \urlstyle{tt}\Url}

\bibitem{Aad:2014kva}
\hrefCMSnoop {} {{ ATLAS} Collaboration, ``{Measurement of the $t\bar{t}$
  production cross-section using $e\mu$ events with $b$-tagged jets in $pp$
  collisions at $\sqrt{s}=7$ and 8 TeV with the ATLAS detector}'',} (2014).
\href{http://www.arXiv.org/abs/1406.5375}{\texttt{ arXiv:1406.5375}}.

\bibitem{Chatrchyan:2013faa}
\hrefCMSnoop {} {{ CMS} Collaboration, ``{Measurement of the \ttbar production
  cross section in the dilepton channel in pp collisions at $\sqrt{s} =
  8$\TeV}'',} \textit{ JHEP} \textbf{ 02} (2014) 024,
  \href{http://dx.doi.org/10.1007/JHEP02(2014)024}{\doi{10.1007/JHEP02(2014)024}},
\href{http://www.arXiv.org/abs/1312.7582}{\texttt{ arXiv:1312.7582}}.

\bibitem{Chatrchyan:2012vs}
\hrefCMSnoop {} {{ CMS} Collaboration, ``{Measurement of the top quark pair
  production cross section in pp collisions at $\sqrt{s} = 7$ TeV in dilepton
  final states containing a $\tau$}'',} \textit{ Phys. Rev. D} \textbf{ 85}
  (2012) 112007,
  \href{http://dx.doi.org/10.1103/PhysRevD.85.112007}{\doi{10.1103/PhysRevD.85.112007}},
\href{http://www.arXiv.org/abs/1203.6810}{\texttt{ arXiv:1203.6810}}.

\bibitem{Gunion:1989we}
J.~F. Gunion, H.~E. Haber, G.~L. Kane, and S.~Dawson, ``The Higgs Hunter's
  Guide''.
\newblock Frontiers in Physics. Addison-Wesley, 1990.

\bibitem{Djouadi:2005gj}
\hrefCMSnoop {} {A.~Djouadi, ``{The anatomy of electro-weak symmetry breaking
  Tome II: The Higgs bosons in the minimal supersymmetric model}'',} \textit{
  Phys. Rept.} \textbf{ 459} (2008) 1,
  \href{http://dx.doi.org/10.1016/j.physrep.2007.10.005}{\doi{10.1016/j.physrep.2007.10.005}},
\href{http://www.arXiv.org/abs/hep-ph/0503173}{\texttt{ arXiv:hep-ph/0503173}}.

\bibitem{Chatrchyan:2012vca}
\hrefCMSnoop {} {{ CMS} Collaboration, ``{Search for a light charged Higgs
  boson in top quark decays in pp collisions at $\sqrt{s}=7$ TeV}'',} \textit{
  JHEP} \textbf{ 07} (2012) 143,
  \href{http://dx.doi.org/10.1007/JHEP07(2012)143}{\doi{10.1007/JHEP07(2012)143}},
\href{http://www.arXiv.org/abs/1205.5736}{\texttt{ arXiv:1205.5736}}.

\bibitem{c:taudil_cdf1}
\hrefCMSnoop {} {{ CDF} Collaboration, ``The $\mu\tau$ and $e\tau$ decays of
  top quark pairs produced in $p\bar{p}$ collisions at $\sqrt{s} =
  1.8$~{TeV}'',} \textit{ Phys. Rev. Lett.} \textbf{ 79} (1997) 3585,
  \href{http://dx.doi.org/10.1103/PhysRevLett.79.3585}{\doi{10.1103/PhysRevLett.79.3585}},
\href{http://www.arXiv.org/abs/hep-ex/9704007}{\texttt{ arXiv:hep-ex/9704007}}.

\bibitem{c:taudil_cdf2}
\hrefCMSnoop {} {{ CDF} Collaboration, ``{A search for $t \to \tau \nu q$ in
  \ttbar production}'',} \textit{ Phys. Lett. B} \textbf{ 639} (2006) 172,
  \href{http://dx.doi.org/10.1016/j.physletb.2006.06.030}{\doi{10.1016/j.physletb.2006.06.030}},
\href{http://www.arXiv.org/abs/hep-ex/0510063}{\texttt{ arXiv:hep-ex/0510063}}.

\bibitem{c:d0taudil}
\hrefCMSnoop {} {{ D0} Collaboration, ``{Measurement of the \ttbar production
  cross section and top quark mass extraction using dilepton events in
  $\mathrm{p\bar{p}}$ collisions}'',} \textit{ Phys. Lett. B} \textbf{ 679}
  (2009) 177,
  \href{http://dx.doi.org/10.1016/j.physletb.2009.07.032}{\doi{10.1016/j.physletb.2009.07.032}},
\href{http://www.arXiv.org/abs/0901.2137}{\texttt{ arXiv:0901.2137}}.

\bibitem{Aaltonen:2014hua}
\hrefCMSnoop {} {{ CDF} Collaboration, ``{Study of top quark production and
  decays involving a tau lepton at CDF and limits on a charged Higgs boson
  contribution}'',} \textit{ Phys. Rev. D} \textbf{ 89} (2014) 091101,
  \href{http://dx.doi.org/10.1103/PhysRevD.89.091101}{\doi{10.1103/PhysRevD.89.091101}},
\href{http://www.arXiv.org/abs/1402.6728}{\texttt{ arXiv:1402.6728}}.

\bibitem{Aad:2012mza}
\hrefCMSnoop {} {{ ATLAS} Collaboration, ``{Measurement of the top quark pair
  cross section with ATLAS in pp collisions at $\sqrt{s} = 7$~TeV using final
  states with an electron or a muon and a hadronically decaying $\tau$
  lepton}'',} \textit{ Phys. Lett. B} \textbf{ 717} (2012) 89,
  \href{http://dx.doi.org/10.1016/j.physletb.2012.09.032}{\doi{10.1016/j.physletb.2012.09.032}},
\href{http://www.arXiv.org/abs/1205.2067}{\texttt{ arXiv:1205.2067}}.

\bibitem{JINST}
\hrefCMSnoop {} {{ CMS} Collaboration, ``{The CMS experiment at the CERN
  LHC}'',} \textit{ JINST} \textbf{ 3} (2008) S08004,
\href{http://dx.doi.org/10.1088/1748-0221/3/08/S08004}{\doi{10.1088/1748-0221/3/08/S08004}}.

\bibitem{Alwall:2011uj}
J.~Alwall\hrefCMSnoop {} { {et~al.}, ``{MadGraph} 5: going beyond'',} \textit{
  JHEP} \textbf{ 06} (2011) 128,
  \href{http://dx.doi.org/10.1007/JHEP06(2011)128}{\doi{10.1007/JHEP06(2011)128}},
\href{http://www.arXiv.org/abs/1106.0522}{\texttt{ arXiv:1106.0522}}.

\bibitem{Alwall:2014hca}
J.~Alwall\hrefCMSnoop {} { {et~al.}, ``The automated computation of tree-level
  and next-to-leading order differential cross sections, and their matching to
  parton shower simulations'',} \textit{ JHEP} \textbf{ 07} (2014) 079,
  \href{http://dx.doi.org/10.1007/JHEP07(2014)079}{\doi{10.1007/JHEP07(2014)079}},
\href{http://www.arXiv.org/abs/1405.0301}{\texttt{ arXiv:1405.0301}}.

\bibitem{Czakon:2013goa}
\hrefCMSnoop {} {M.~Czakon, P.~Fiedler, and A.~Mitov, ``{The total top quark
  pair production cross section at hadron colliders through
  O($\alpha_S^4$)}'',} \textit{ Phys. Rev. Lett.} \textbf{ 110} (2013) 252004,
  \href{http://dx.doi.org/10.1103/PhysRevLett.110.252004}{\doi{10.1103/PhysRevLett.110.252004}},
\href{http://www.arXiv.org/abs/1303.6254}{\texttt{ arXiv:1303.6254}}.

\bibitem{Czakon:2011xx}
\hrefCMSnoop {} {M.~Czakon and A.~Mitov, ``{Top++: A program for the
  calculation of the top pair cross-section at hadron colliders}'',} \textit{
  Comput. Phys. Commun.} \textbf{ 185} (2014) 2930,
  \href{http://dx.doi.org/10.1016/j.cpc.2014.06.021}{\doi{10.1016/j.cpc.2014.06.021}},
\href{http://www.arXiv.org/abs/1112.5675}{\texttt{ arXiv:1112.5675}}.

\bibitem{Botje:2011sn}
M.~Botje\hrefCMSnoop {} { {et~al.}, ``{The PDF4LHC Working Group Interim
  Recommendations}'',} (2011).
\href{http://www.arXiv.org/abs/1101.0538}{\texttt{ arXiv:1101.0538}}.

\bibitem{Gao:2013xoa}
J.~Gao\hrefCMSnoop {} { {et~al.}, ``{CT10 next-to-next-to-leading order global
  analysis of QCD}'',} \textit{ Phys. Rev. D} \textbf{ 89} (2014) 033009,
  \href{http://dx.doi.org/10.1103/PhysRevD.89.033009}{\doi{10.1103/PhysRevD.89.033009}},
\href{http://www.arXiv.org/abs/1302.6246}{\texttt{ arXiv:1302.6246}}.

\bibitem{Ball:2012cx}
\hrefCMSnoop {} {{ NNPDF} Collaboration, ``{Parton distributions with LHC
  data}'',} \textit{ Nucl. Phys. B} \textbf{ 867} (2013) 244,
  \href{http://dx.doi.org/10.1016/j.nuclphysb.2012.10.003}{\doi{10.1016/j.nuclphysb.2012.10.003}},
\href{http://www.arXiv.org/abs/1207.1303}{\texttt{ arXiv:1207.1303}}.

\bibitem{pythia}
\hrefCMSnoop {} {T.~Sj{\"o}strand, S.~Mrenna, and P.~Z. Skands, ``{PYTHIA} 6.4
  physics and manual'',} \textit{ JHEP} \textbf{ 05} (2006) 026,
  \href{http://dx.doi.org/10.1088/1126-6708/2006/05/026}{\doi{10.1088/1126-6708/2006/05/026}},
  \href{http://www.arXiv.org/abs/hep-ph/0603175}{\texttt{
  arXiv:hep-ph/0603175}}.

\bibitem{Alwall:2007fs}
J.~Alwall\hrefCMSnoop {} { {et~al.}, ``Comparative study of various algorithms
  for the merging of parton showers and matrix elements in hadronic
  collisions'',} \textit{ Eur. Phys. J. C} \textbf{ 53} (2008) 473,
  \href{http://dx.doi.org/10.1140/epjc/s10052-007-0490-5}{\doi{10.1140/epjc/s10052-007-0490-5}},
\href{http://www.arXiv.org/abs/0706.2569}{\texttt{ arXiv:0706.2569}}.

\bibitem{tauola2}
P.~Golonka\hrefCMSnoop {} { {et~al.}, ``{The tauola-photos-F environment for
  the TAUOLA and PHOTOS packages, release II}'',} \textit{ Comput. Phys.
  Commun.} \textbf{ 174} (2006) 818,
  \href{http://dx.doi.org/10.1016/j.cpc.2005.12.018}{\doi{10.1016/j.cpc.2005.12.018}},
\href{http://www.arXiv.org/abs/hep-ph/0312240}{\texttt{ arXiv:hep-ph/0312240}}.

\bibitem{Artoisenet:2012st}
\hrefCMSnoop {} {P.~Artoisenet, R.~Frederix, O.~Mattelaer, and R.~Rietkerk,
  ``{Automatic spin-entangled decays of heavy resonances in Monte Carlo
  simulations}'',} \textit{ JHEP} \textbf{ 03} (2013) 015,
  \href{http://dx.doi.org/10.1007/JHEP03(2013)015}{\doi{10.1007/JHEP03(2013)015}},
\href{http://www.arXiv.org/abs/1212.3460}{\texttt{ arXiv:1212.3460}}.

\bibitem{Nason:2004rx}
\hrefCMSnoop {} {P.~Nason, ``{A New method for combining NLO QCD with shower
  Monte Carlo algorithms}'',} \textit{ JHEP} \textbf{ 11} (2004) 040,
  \href{http://dx.doi.org/10.1088/1126-6708/2004/11/040}{\doi{10.1088/1126-6708/2004/11/040}},
\href{http://www.arXiv.org/abs/hep-ph/0409146}{\texttt{ arXiv:hep-ph/0409146}}.

\bibitem{powheg}
\hrefCMSnoop {} {S.~Frixione, P.~Nason, and C.~Oleari, ``{Matching NLO QCD
  computations with parton shower simulations: the POWHEG method}'',} \textit{
  JHEP} \textbf{ 11} (2007) 070,
  \href{http://dx.doi.org/10.1088/1126-6708/2007/11/070}{\doi{10.1088/1126-6708/2007/11/070}},
  \href{http://www.arXiv.org/abs/0709.2092}{\texttt{ arXiv:0709.2092}}.

\bibitem{Alioli:2010xd}
\hrefCMSnoop {} {S.~Alioli, P.~Nason, C.~Oleari, and E.~Re, ``{A general
  framework for implementing NLO calculations in shower Monte Carlo programs:
  the POWHEG BOX}'',} \textit{ JHEP} \textbf{ 06} (2010) 043,
  \href{http://dx.doi.org/10.1007/JHEP06(2010)043}{\doi{10.1007/JHEP06(2010)043}},
\href{http://www.arXiv.org/abs/1002.2581}{\texttt{ arXiv:1002.2581}}.

\bibitem{Alioli:2009je}
\hrefCMSnoop {} {S.~Alioli, P.~Nason, C.~Oleari, and E.~Re, ``{NLO single-top
  production matched with shower in POWHEG: $s$- and $t$-channel
  contributions}'',} \textit{ JHEP} \textbf{ 09} (2009) 111,
  \href{http://dx.doi.org/10.1088/1126-6708/2009/09/111}{\doi{10.1088/1126-6708/2009/09/111}},
  \href{http://www.arXiv.org/abs/0907.4076}{\texttt{ arXiv:0907.4076}}.
[Erratum: \DOI{10.1007/JHEP02(2010)011}].

\bibitem{Re:2010bp}
\hrefCMSnoop {} {E.~Re, ``{Single-top Wt-channel production matched with parton
  showers using the POWHEG method}'',} \textit{ Eur. Phys. J. C} \textbf{ 71}
  (2011) 1547,
  \href{http://dx.doi.org/10.1140/epjc/s10052-011-1547-z}{\doi{10.1140/epjc/s10052-011-1547-z}},
\href{http://www.arXiv.org/abs/1009.2450}{\texttt{ arXiv:1009.2450}}.

\bibitem{Chatrchyan:2011id}
\hrefCMSnoop {} {{ CMS} Collaboration, ``{Measurement of the underlying event
  activity at the LHC with $\sqrt{s}= 7$~TeV and comparison with
  $\sqrt{s}=0.9$~TeV}'',} \textit{ JHEP} \textbf{ 11} (2011) 109,
  \href{http://dx.doi.org/10.1007/JHEP09(2011)109}{\doi{10.1007/JHEP09(2011)109}},
\href{http://www.arXiv.org/abs/1107.0330}{\texttt{ arXiv:1107.0330}}.

\bibitem{pdfset}
P.~M. Nadolsky\hrefCMSnoop {} { {et~al.}, ``{Implications of CTEQ global
  analysis for collider observables}'',} \textit{ Phys. Rev. D} \textbf{ 78}
  (2008) 013004,
  \href{http://dx.doi.org/10.1103/PhysRevD.78.013004}{\doi{10.1103/PhysRevD.78.013004}},
\href{http://www.arXiv.org/abs/0802.0007}{\texttt{ arXiv:0802.0007}}.

\bibitem{Agostinelli:2002hh}
\hrefCMSnoop {} {{ GEANT4} Collaboration, ``{GEANT4}---a simulation toolkit'',}
  \textit{ Nucl. Instrum. Meth. A} \textbf{ 506} (2003) 250,
\href{http://dx.doi.org/10.1016/S0168-9002(03)01368-8}{\doi{10.1016/S0168-9002(03)01368-8}}.

\bibitem{Allison:2006ve}
\hrefCMSnoop {} {J.~Allison {et~al.}, ``{GEANT4 developments and
  applications}'',} \textit{ IEEE Trans. Nucl. Sci.} \textbf{ 53} (2006) 270,
\href{http://dx.doi.org/10.1109/TNS.2006.869826}{\doi{10.1109/TNS.2006.869826}}.

\bibitem{CMS-PAS-PFT-09-001}
\href {http://cdsweb.cern.ch/record/1194487} {{ CMS} Collaboration, ``Particle
  flow event reconstruction in {CMS} and performance for jets, taus, and
  {\MET}'',} CMS Physics Analysis Summary CMS-PAS-PFT-09-001, 2009.

\bibitem{CMS-PAS-PFT-10-001}
\href {http://cdsweb.cern.ch/record/1247373} {{ CMS} Collaboration,
  ``Commissioning of the particle flow event reconstruction with the first
  {LHC} collisions recorded in the {CMS} detector'',} CMS Physics Analysis
  Summary CMS-PAS-PFT-10-001, 2010.

\bibitem{CMS-PAS-EGM-10-004}
\href {http://cdsweb.cern.ch/record/1299116} {{ CMS} Collaboration, ``{Electron
  reconstruction and Identification at $\sqrt{s} = 7$~TeV}'',} CMS Physics
  Analysis Summary CMS-PAS-EGM-10-004, 2010.

\bibitem{Chatrchyan:2012xi}
\hrefCMSnoop {} {{ CMS} Collaboration, ``{Performance of CMS muon
  reconstruction in pp collision events at $\sqrt{s}=7$~TeV}'',} \textit{
  JINST} \textbf{ 7} (2012) P10002,
  \href{http://dx.doi.org/10.1088/1748-0221/7/10/P10002}{\doi{10.1088/1748-0221/7/10/P10002}},
\href{http://www.arXiv.org/abs/1206.4071}{\texttt{ arXiv:1206.4071}}.

\bibitem{antikt}
\hrefCMSnoop {} {M.~Cacciari, G.~P. Salam, and G.~Soyez, ``{The anti-$k_t$ jet
  clustering algorithm}'',} \textit{ JHEP} \textbf{ 04} (2008) 063,
  \href{http://dx.doi.org/10.1088/1126-6708/2008/04/063}{\doi{10.1088/1126-6708/2008/04/063}},
\href{http://www.arXiv.org/abs/0802.1189}{\texttt{ arXiv:0802.1189}}.

\bibitem{jme-10-011}
\hrefCMSnoop {} {{ CMS} Collaboration, ``{Determination of jet energy
  calibration and transverse momentum resolution in CMS}'',} \textit{ JINST}
  \textbf{ 6} (2011) P11002,
  \href{http://dx.doi.org/10.1088/1748-0221/6/11/P11002}{\doi{10.1088/1748-0221/6/11/P11002}},
\href{http://www.arXiv.org/abs/1107.4277}{\texttt{ arXiv:1107.4277}}.

\bibitem{Chatrchyan:2012zz}
\hrefCMSnoop {} {{ CMS} Collaboration, ``{Performance of tau lepton
  reconstruction and identification in CMS}'',} \textit{ JINST} \textbf{ 7}
  (2012) P01001,
  \href{http://dx.doi.org/10.1088/1748-0221/7/01/P01001}{\doi{10.1088/1748-0221/7/01/P01001}},
\href{http://www.arXiv.org/abs/1109.6034}{\texttt{ arXiv:1109.6034}}.

\bibitem{Chatrchyan:2014nva}
\hrefCMSnoop {} {{ CMS} Collaboration, ``{Evidence for the 125 GeV Higgs boson
  decaying to a pair of $\tau$ leptons}'',} \textit{ JHEP} \textbf{ 05} (2014)
  104,
  \href{http://dx.doi.org/10.1007/JHEP05(2014)104}{\doi{10.1007/JHEP05(2014)104}},
\href{http://www.arXiv.org/abs/1401.5041}{\texttt{ arXiv:1401.5041}}.

\bibitem{Chatrchyan:2012jua}
\hrefCMSnoop {} {{ CMS} Collaboration, ``{Identification of b-quark jets with
  the CMS experiment}'',} \textit{ JINST} \textbf{ 8} (2013) P04013,
  \href{http://dx.doi.org/10.1088/1748-0221/8/04/P04013}{\doi{10.1088/1748-0221/8/04/P04013}},
\href{http://www.arXiv.org/abs/1211.4462}{\texttt{ arXiv:1211.4462}}.

\bibitem{Chatrchyan:2012ea}
\hrefCMSnoop {} {{ CMS} Collaboration, ``{Measurement of the top-quark mass in
  \ttbar events with dilepton final states in pp collisions at
  $\sqrt{s}=7$\TeV}'',} \textit{ Eur. Phys. J. C} \textbf{ 72} (2012) 2202,
  \href{http://dx.doi.org/10.1140/epjc/s10052-012-2202-z}{\doi{10.1140/epjc/s10052-012-2202-z}},
\href{http://www.arXiv.org/abs/1209.2393}{\texttt{ arXiv:1209.2393}}.

\bibitem{Chatrchyan:2011nb}
\hrefCMSnoop {} {{ CMS} Collaboration, ``{Measurement of the \ttbar production
  cross section and the top quark mass in the dilepton channel in pp collisions
  at $\sqrt{s}=7$ TeV}'',} \textit{ JHEP} \textbf{ 07} (2011) 049,
  \href{http://dx.doi.org/10.1007/JHEP07(2011)049}{\doi{10.1007/JHEP07(2011)049}},
\href{http://www.arXiv.org/abs/1105.5661}{\texttt{ arXiv:1105.5661}}.

\bibitem{CMS-PAS-LUM-13-001}
\href {http://cds.cern.ch/record/1598864?ln=en} {{ CMS} Collaboration, ``CMS
  luminosity based on pixel cluster counting - Summer 2013 Update'',} CMS
  Physics Analysis Summary CMS-PAS-LUM-13-001, 2013.

\bibitem{Alioli:2011as}
\hrefCMSnoop {} {S.~Alioli, S.-O. Moch, and P.~Uwer, ``{Hadronic top-quark
  pair-production with one jet and parton showering}'',} \textit{ JHEP}
  \textbf{ 01} (2012) 137,
  \href{http://dx.doi.org/10.1007/JHEP01(2012)137}{\doi{10.1007/JHEP01(2012)137}},
\href{http://www.arXiv.org/abs/1110.5251}{\texttt{ arXiv:1110.5251}}.

\bibitem{blue}
\hrefCMSnoop {} {L.~Lyons, D.~Gibaut, and P.~Clifford, ``How to combine
  correlated estimates of a single physical quantity'',} \textit{ Nucl.
  Instrum. Meth. A} \textbf{ 270} (1988) 110,
\href{http://dx.doi.org/10.1016/0168-9002(88)90018-6}{\doi{10.1016/0168-9002(88)90018-6}}.

\bibitem{ATLAS:2014wva}
\hrefCMSnoop {} {{ATLAS, CDF, CMS, and D0 collaborations}, ``{First combination
  of Tevatron and LHC measurements of the top quark mass}'',} (2014).
\href{http://www.arXiv.org/abs/1403.4427}{\texttt{ arXiv:1403.4427}}.

\end{thebibliography}\endgroup

\cleardoublepage \appendix\section{The CMS Collaboration \label{app:collab}}\begin{sloppypar}\hyphenpenalty=5000\widowpenalty=500\clubpenalty=5000\textbf{Yerevan Physics Institute,  Yerevan,  Armenia}\\*[0pt]
V.~Khachatryan, A.M.~Sirunyan, A.~Tumasyan
\vskip\cmsinstskip
\textbf{Institut f\"{u}r Hochenergiephysik der OeAW,  Wien,  Austria}\\*[0pt]
W.~Adam, T.~Bergauer, M.~Dragicevic, J.~Er\"{o}, C.~Fabjan\cmsAuthorMark{1}, M.~Friedl, R.~Fr\"{u}hwirth\cmsAuthorMark{1}, V.M.~Ghete, C.~Hartl, N.~H\"{o}rmann, J.~Hrubec, M.~Jeitler\cmsAuthorMark{1}, W.~Kiesenhofer, V.~Kn\"{u}nz, M.~Krammer\cmsAuthorMark{1}, I.~Kr\"{a}tschmer, D.~Liko, I.~Mikulec, D.~Rabady\cmsAuthorMark{2}, B.~Rahbaran, H.~Rohringer, R.~Sch\"{o}fbeck, J.~Strauss, A.~Taurok, W.~Treberer-Treberspurg, W.~Waltenberger, C.-E.~Wulz\cmsAuthorMark{1}
% ...
\vskip\cmsinstskip
% ...
1:~~Also at Vienna University of Technology, Vienna, Austria\\
2:~~Also at CERN, European Organization for Nuclear Research, Geneva, Switzerland\\
% ...

\end{sloppypar}
\end{document}